\documentclass[%
 aip,
 jmp,
 amsmath,amssymb,
 reprint,%
]{revtex4-1}
\usepackage{mathptmx}
\usepackage{mathrsfs}
\usepackage{graphicx}
\usepackage{dcolumn}
\usepackage{bm}
\usepackage{caption}
\usepackage{CJK}
\usepackage{etoolbox}
\usepackage{float}
\usepackage{amstext}
\makeatletter
\def\@email#1#2{%
 \endgroup
 \patchcmd{\titleblock@produce}
  {\frontmatter@RRAPformat}
  {\frontmatter@RRAPformat{\produce@RRAP{*#1\href{mailto:#2}{#2}}}\frontmatter@RRAPformat}
  {}{}
}%
\makeatother
\begin{document}


\title[]{Competition Between Thermophoretic Separation and Thermal
Convective Mixing}
\author{Yu Lu
\begin{CJK}{UTF8}{gbsn}
(陆钰)
\end{CJK}
}
\affiliation{ 
School of Mechanical Engineering,
Nantong University,
Nantong 226019, China
}
\affiliation{ 
Shanghai Institute of Applied Mathematics and Mechanics, School of Mechanics and Engineering Science, Shanghai Key Laboratory of Mechanics in Energy Engineering,
Shanghai Frontier Science Center of Mechanoinformatics, Shanghai University,
Shanghai 200072, China
}
\author{Guo-Hui Hu
\begin{CJK}{UTF8}{gbsn}
(胡国辉)
\end{CJK}
$^*$}%
\email{ghhu@staff.shu.edu.cn}
\affiliation{ 
Shanghai Institute of Applied Mathematics and Mechanics, School of Mechanics and Engineering Science, Shanghai Key Laboratory of Mechanics in Energy Engineering,
Shanghai Frontier Science Center of Mechanoinformatics, Shanghai University,
Shanghai 200072, China
}

\date{May 24, 2025}

\begin{abstract}
In binary fluid systems under temperature differences, thermophoretic separation and thermal convective mixing are two key mechanisms that affect the processes of transport. The thermophoretic effect, also known as the Soret effect, describes the migration behavior of molecules in the fluid with temperature gradient. Thermophilic and thermophobic molecules tend to migrate to regions of higher and lower temperature.
Thermal convective mixing is triggered by a Rayleigh-B{\'e}nard-type hydrodynamics instability as a macroscopic flow caused by buoyancy induced by temperature differences, promoting the mixing of different components within the fluid.
There is a competition between the separation caused by thermophoresis and the mixing produced by convection. 
A dimensionless number, namely Soret-Rayleigh number $S_R$, which is the ratio of dimensionless Soret factor $S_T$ and Rayleigh numbers $Ra$, is introduced to quantitatively describe this competition.
The impact of different $S_R$ and Rayleigh numbers $Ra$ on the concentration field is studied by energy-conserving dissipative particle dynamics (eDPD).
The results indicate that the components separation exhibits a non-monotonic variation with $Ra$ increase. 
For a specific $S_R$, there exists an optimal $Ra$ that achieves the lowest separation at long time scales. 
This result will have potential implications for the design and application of micro-mixers.
\end{abstract}

\maketitle
\section{Introduction}
The thermal convection of binary fluids holds significant value in both theoretical\cite{hu2021new,khandelwal2021instabilities} and practical fields\cite{knobloch1980convection,tait1989compositional,schopf1993convection,rosner2005calculation,rosner2007soret}.
Non-uniform buoyancy forces induced by the temperature difference, leading to the hydrodynamic instability known as Rayleigh-B{\'e}nard (RB) convection\cite{niemela2000turbulent,grossmann2000scaling,ecke2023turbulent,shishkina2016momentum}.
Different from pure fluids, in flow fields with temperature gradients, different components within binary fluids exhibit separation along or counter to the temperature gradient direction. This phenomenon is termed the Soret effect\cite{huang2010isotope,dominguez2011soret,Duhr2006Thermophoretic,wienken2010protein,Reichl2014Why,Rudani2024Deciphering,Lu2024single} (also referred to as thermophoresis). The coupling of these two phenomena generates complex and intriguing flow patterns and separation behaviors.

For binary fluid mixtures where the two components exhibit similar densities, variations in their local concentration negligibly affect fluid density and buoyancy forces. In such cases, the concentration field acts as a passive scalar, as it does not significantly influence the flow dynamics.
Conversely, in systems such as water-ethanol mixtures\cite{moses1986competing}, $^3$He-$^4$He mixtures\cite{ahlers1986convection}, or nanoparticles suspension\cite{kim2006analysis}, local concentration changes markedly vary fluid density and buoyancy, thereby directly affect the flow. Here, the concentration field serves as an active scalar due to its coupling with hydrodynamic instabilities.
The most critical dimensionless parameter in this context is the separation ratio ${\chi}$, which represents the ratio of the buoyancy contributions from the concentration gradient to those from the temperature gradient. When the separation ratio ${\chi}\neq0$, the concentration acts as an active scalar. Conversely, when the separation ratio ${\chi}\approx 0$ is zero or negligible, the concentration becomes a passive scalar.
In recent years, a series of stability analysis had been made for different separation ratio\cite{hu2021new,khandelwal2021instabilities}.

Although many studies have examined binary fluid convection at different separation ratios\cite{moses1986competing,ahlers1986convection,kim2006analysis}, One obvious yet rarely mentioned problem remains. 
Convection creates large-scale flow structure that mix fluid components, while the Soret effect drives them apart through thermal diffusion. This competition happens regardless of the separation ratio.

The competition between convective mixing and Soret-driven separation is governed by two dimensionless parameters.
The Rayleigh number $Ra$, which characterizes buoyancy and governs convection and mixing is proportional to the temperature difference $\theta^{*}$.
If we rescale the Soret coefficient $S_T^{*}\sim 1/K$ by temperature difference $\theta^{*}$, we get the dimensionless Soret coefficient $S_T=S_T^{*} \theta^{*}$, which governs separation of two components. 
Notably, both of these dimensionless parameters are proportional to an externally controllable quantity, the temperature difference $ \theta^{*}$. 
Remarkably, their ratio forms a unique dimensionless number determined solely by system geometry and fluid properties.

Naturally, a new dimensionless number emerges to characterize the competition between convective mixing and Soret separation. We tentatively refer to it as the Soret-Rayleigh number $S_R=S_T/Ra$. The $S_R$ is independent of the external temperature difference $ \theta^{*}$, meaning that the $S_R$ number and the temperature difference can describe the RB convection in binary fluid mixtures with the Soret effect from two distinct and independent perspectives. 
Then, two crucial questions have to be examined: (1) How to determine the magnitude of the $S_R$ number, or in other words, what do different values of the $S_R$ number signify? (2) How do the $S_R$ number and the temperature difference $\theta^{*}$ (or $Ra$) influence the physical fields in binary fuilds, particularly the concentration field?

To address the aforementioned questions, we aim to investigate the problem using a simplest model. In this model, the influence of concentration on the flow is neglected, rendering the concentration a passive scalar. Furthermore, we focus solely on steady-state results, i.e., $\partial C / \partial t=0$. Under these conditions, after prolonged diffusion and convective mixing, the concentration fields under different parameters become difficult to distinguish. Instead of the commonly used mixing index, we define a separation index to characterize the degree of separation in the binary fluid mixture.

In this paper, we employ a mesoscopic simulation approach to investigate a simplified model, aiming to analyze the effects of thermal convection and the Soret effect on the concentration field in binary fluid mixtures. We introduce a new dimensionless number, $S_R$, to characterize the competition between convective mixing and Soret separation. The relationship between the separation index and the temperature difference is analyzed for various $S_R$ numbers. Additionally, we propose a scaling theory for the separation index as a function of the dimensionless parameters.

\section{Method and Model}
\subsection{Governing equations of eDPD}
The coarse-grained molecular dynamics method Dissipative Particle Dynamics (DPD) was first introduced by Hoogerbrugge and Koelman\cite{hoogerbrugge1992simulating}, which has been successfully applied to many biological and chemical systems\cite{camerin2018modelling,nikolov2018mesoscale,chen2018dissipative,Lu2021potential,Lu2022Double,lu2023linear,pan2009rheology,li2015mesoscale,tang2016non}.

In eDPD method, the simulation system contains a collection of finite-size particles with mass $m_{i}$ and heat capacity $C_{vi}$ for the $i$th particle.
The governing equation of particle motion is the conservation of momentum and energy:
\begin{equation}\label{equ:NSL}
\frac{d \vec{r}_{i}}{d t}=\vec{v}_{i} ; \ \ \  m_{i} \frac{d^{2} \vec{r}_{i}}{d t^{2}}=\vec{F}_{i}=\sum_{j} \vec{F}_{i j},
\end{equation}
\begin{equation}\label{equ:Energy}
    C_{vi}\frac{\partial T_i}{\partial t}=q_i=\sum_{i \neq j}q_{ij},
\end{equation}
where $\vec{r}_{i}, \vec{v}_{i}$ and $T_i$ are the vector of $i$th particle's position, velocity,and temperature, 
$\vec{r}_{i j}=\vec{r}_{i}-\vec{r}_{j}$. The magnitude of the vector ${r}_{i j}=|\vec{r}_{i j}|$ is the distance between particle $j$ and particle $i$.

$\vec{F}_{i}$ is the total force acting on the particle $i$. $\vec{F}_{i j}$ is the force exerted on particle $i$ by particle $j$, which consists of three parts:
\begin{equation}\label{equ:TF}
\vec{F}_{i j}=\vec{F}_{i j}^{C}+\vec{F}_{i j}^{D}+\vec{F}_{i j}^{R}.
\end{equation}
The conservative force is given by
\begin{equation}\label{equ:FC}
{\vec{F}}_{ij}^C=a_{ij}w_C(r_{ij}){\vec{e}}_{ij}
\end{equation}
In traditional DPD, $a_{ij}$ is the maximum of conservative force,  ${\vec{e}}_{ij}=\vec{r}_{ij}/{r}_{ij}$ is a unit vector,
and the weight function is given by
\begin{equation}\label{equ:WC}
w_C(r_{ij})=
\left\{
\begin{aligned}
&1-\frac{r_{ij}}{R_{CC}}&,\ r_{ij}<R_{CC}
\\
&0						  &,\ r_{ij}> R_{CC}
\end{aligned}
\right.,
\end{equation}
where $R_{CC}$ is the cut-off radius of conservative force.
The dissipative force is 
\begin{equation}\label{equ:FDc}
{\vec{F}}_{ij}^{D}=-\gamma_{ij}w_D(r_{ij})({\vec{e}}_{ij}\cdot{\vec{v}}_{ij}){\vec{e}}_{ij};
\end{equation}

where ${\vec{v}}_{ij}={\vec{v}}_i-{\vec{v}}_j$ is relative velocity and $w_D(r_{ij})$ is the weight function for dissipative force.
$\gamma_{ij}$ is dissipative coefficients.

The random force is defined by
\begin{equation}\label{equ:FR}
{\vec{F}}_{ij}^Rdt=w_R(r_{ij})\sigma_{ij}dW_{ij}]{\vec{e}}_{ij}
\end{equation}
where $dt$ is the time step and $W_{ij}$ is the weiner process, $\sigma_{ij}$ is the parameters in random force and $w_R(r_{ij})$ is the weight function of random force, which  follows the fluctuation-dissipative theorem
\begin{equation}\label{equ:FDT1}
\sigma_{ij}^2=\frac{4k_B\gamma_{ij}T_iT_j}{T_i+T_j} 
\end{equation}  
\begin{equation}\label{equ:FDT2}
w_D(r_{ij})=w^2_R(r_{ij}))=(1-\frac{r_{ij}}{R_{CD}})^{s},
\end{equation}  
where $k_B$ is the Boltzmann constant,
$R_{CD}$ is the cut-off radius for dissipative and random force,
and $s$ is the exponent of weight function\cite{Fan2006Simulating}.

$\vec{q}_{i}$ is the total heat flux that acts on the particle $i$. $\vec{q}_{i j}$ is the heat flux exerted on particle $i$ by particle $j$, which consists of three parts:
\begin{equation}\label{equ:qij3Part}
q_{ij}=q^{C}_{ij}+q^{V}_{ij}+q^{R}_{ij}
\end{equation}
in which $q_{ij}^C$ and $q^{R}_{ij}$ is the collisional and random heat flux, given by
\begin{equation}\label{equ:qc}
\begin{aligned}
&q^{C}_{ij}=k_{ij}w_{CT}(r_{ij})(\frac{1}{T_i}-\frac{1}{T_j}),\\
&q_{ij}^{R}dt=\beta_{ij}w_{RT}(r_{ij})dW_{ij}^e.\\ 
\end{aligned}
\end{equation}
where $k_{ij}=C_{vi}C_{vj}\kappa T_{ij}^2/k_B$ in which $\kappa$ is the mesoscale heat friction coefficient\cite{Li2014Energy}, and $\beta_{ij}^2=k_{ij}k_B$.
The weight function $w_{CT}(r_{ij})$ and $w_{RT}(r_{ij})$ are given as $w_{CT}(r_{ij})=w_{RT}^2(r_{ij})=1-r_{ij}/R_{CT}$ with the cut-off radius $R_{CT}$.
The viscous heat flux is given by
\begin{equation}\label{equ:qvc}
\begin{aligned}
  &(C_{vi}+C_{vj}) q^{V}_{ij}dt
  = w_D(r_{ij})[(\vec{e_{ij}} \cdot \vec{v_{ij}})^2-\frac{\sigma_{ij}}{m}]dt \\
  &-\sigma_{ij}w_R(r_{ij})(\vec{e_{ij}} \cdot \vec{v_{ij}})dW_{ij},  \\
\end{aligned}
\end{equation}
which is the work done by the dissipative and random force.

\subsection{Boussinesq approximation and Soret effect in eDPD}
Based on the Boussinesq approximation, the buoyancy force is applied on the each DPD particle\cite{lu2024simulation}, which is given by
\begin{equation}
    \vec{F_i}^b=g\beta(T_i-T_0)
\end{equation}
in which $g$ and $\beta$ is gravitational acceleration and coefficient of thermal
expansion in DPD units. 

To model the Soret effect in a DPD system, the key problem is to apply thermophoretic force $F^T$ on a DPD particle. A modified conservative force\cite{Lu2024single} is utilized:
\begin{equation}\label{equ:SoretForce}
\begin{aligned}
   &{\vec{F}}_{ij}^{C}=a_{ij}w_C(r_{ij}){\vec{e}}_{ij}+{\vec{F}}_{ij}^{CT} ,\\
   &{\vec{F}}_{ij}^{CT}=-a^T_{ij}(T_i-T_j)w_{CT}(r_{ij})  \vec{e_{ij}}.\\
\end{aligned}
\end{equation}

in which ${\vec{F}}_{ij}^{CT}$ provides the in-homogeneous interaction between thermophilic and thermophobic particles in the temperature gradient.
To satisfy Newton's third law ($F_{ij}^{CT}=-F_{ji}^{CT}$), the parameter must be set to $a_{ij}^{T}=-a_{ji}^T$. For a positive Soret coefficient of the particle $i$, $a^T_{ij}$ should be positive, and when $T_{i}>T_{j}$ the particle with higher temperature will move to the region with lower temperature.

\subsection{Map to real units and dimensionless number}
For the fluid with velocity $v^{*}$ and temperature $T^{*}$ between two walls with high temperature $T_h^{*}$ in the bottom and low temperature $T_c^{*}$ in the top.

In steady state, the Boussinessq equation is
\begin{equation}
v^{*}\frac{dv^{*}}{dx^{*}}=\mu^{*}\frac{d^2v^{*}}{dx^{*2}}+g^{*}\beta^{*}(T^{*}-T_0^{*}),
\end{equation}
in which $T_0^{*}=(T_h^{*}+T_c^{*})/2$, $\mu^{*}$ is the kinetic viscosity, $g^{*}$ and $\beta^{*}$ is gravitational acceleration and
coefficient of thermal expansion.

We consider a binary system with volume fraction $\phi_A=\phi$ and $\phi_B=1-\phi$.
The flux of concentration $J_c^{*}$\cite{leroyer2011soret} and temperature $J_T^{*}$ is
\begin{equation}
\begin{aligned}
    &J_c^{*}/\rho^{*}=v^{*}\phi-D_0^{*}\frac{d\phi}{dx^{*}}-\phi(1-\phi)D_0^{*}S_T^{*}\frac{dT^{*}}{dx^{*}}=0;\\
    &J_T^{*}/\rho^{*}=v^{*}T^{*}-\alpha^{*}\frac{dT^{*}}{dx^{*}}=0,
    \end{aligned}
\end{equation}
in which, $S_T^{*}$ is the Soret factor, $D_0^{*}$ is the self diffusivity and $\alpha^{*}$ is the thermal diffusivity.

Using the following dimensionless variables
\begin{equation}\label{equ:Dless}
\begin{aligned}
    &x=x^{*}/L;\
    v=v^{*}\tau_{\alpha}/L;\
    \tau_{\alpha}=L^2/\alpha^{*};  \\
    &T=\frac{T^{*}-T_0^{*}}{\theta^{*}};  \
    \theta^{*}=T^{*}_h-T^{*}_c\\
\end{aligned}
\end{equation}
in which $\tau_{\alpha}=L^2/\alpha^{*}$ is the characteristic time of thermal conduction.

The other important characteristic time is
\begin{equation}\label{equ:Ctime}
\begin{aligned}
&\tau_d=L^2/D_0^{*}; \ \tau_{\mu}=L^2/\mu^{*};\\
&\tau_s=L^2/D_0^{*}S_T^{*}\theta^{*}; \ \tau_b^2=L/g^{*}\beta^{*}\theta^{*}.\\
\end{aligned}
\end{equation}
Characteristic times for diffusion, viscosity, Soret effect and buoyancy, respectively

The dimensionless equations are written as
\begin{equation}\label{equ:Dless_Be}
\begin{aligned}
&v\frac{dv}{dx}=P_r\frac{d^2v}{dx}+RaP_rT \\
&v\phi-L_e[\frac{d\phi}{dx}-\phi(1-\phi)S_T  \frac{dT}{dx}]=0 \\
&vT-\frac{dT}{dx}=0  \\
\end{aligned}
\end{equation}
in which $P_r={\tau_{\alpha}}/{\tau_{\mu}}$ is Prandtl number, $Ra=\tau_{\alpha}\tau_{\mu}/\tau_{b}^2$ is Rayleigh number, $L_e={\tau_{\alpha}}/{\tau_d}$ is Lewis number, $S_T=\tau_d/\tau_s$ is the dimensionless Soret number.

The Rayleigh number $Ra$ and $S_T$ is proportional to the temperature difference between top and bottom walls $\theta^{*}$. When not changing the physical properties of the fluid in the mixer, for different $\theta^{*}$ the Rayleigh number $Ra$ should be proportional to $S_T$.

Thus, we can define a new dimensionless number 
\begin{equation}\label{equ:S_R}
S_{R}=S_T/Ra=\frac{S_T^{*}\mu^{*}\alpha^{*}}{g^{*}\beta^{*}L^3}\sim {[\theta^{*}]}^0,
\end{equation}
named the Soret-Rayleigh number, which characterizes the competition between thermophoretic separation and thermal convective mixing.

The Soret-Rayleigh number $S_R$ is exclusively determined by the thermophysical properties of the fluid and the characteristic length of the system. When employing typical parameters for water under standard temperature and pressure conditions, with the characterized length $L=10^{4}\mu m$ and Soret coefficients $S_T^{*} \in [10^{-2},10^{-1}] K^{-1}$, the calculated Soret-Rayleigh number $S_R \in [10^{-6} , 10^{-5}]$. The increasing $S_R$ correlated to reduce the system length scale and the larger hydrodynamics radius of molecule.

This paper investigates the Soret separation and convective mixing of binary fluid with different Soret-Rayleigh number and temperature difference. To be conveniently, we define the Soret-Rayleigh number by DPD unit, which is given by 
\begin{equation}
\begin{aligned}
     & S_{R,d}=a^T/g\beta \\
     & S_R=S_{R,d} \frac{\mu\alpha}{H^3}
\end{aligned}
\end{equation}
in which $H=20$ is the height of simulation box, $\mu=0.61$ and $\alpha=0.136$\cite{Lu2024single} is the kinetic viscosity and the thermal
diffusivity in DPD unit.
Considering that the magnitude of $S_R$ number range from $10^{-6}$ to $10^{-5}$, Soret-Rayleigh number by DPD unit $S_{R,d}$ range from $0.1$ to about $0.5$. 

For a certain Soret-Rayleigh number, the temperature difference can be characterized by Rayleigh number.
The Rayleigh number can be calculated by using the physical quantities in DPD units, which is given by
\begin{equation}
    Ra=\frac{g\beta \theta H^3}{\mu\alpha}
\end{equation}
in which $\theta$ is the temperature in DPD unit.
The Rayleigh number is set at approximately $10^4$ to $10^5$, corresponding to a temperature difference $\theta^{*} \in [1,10]K$ in real units and $g\beta\theta \in [0.1,1]$ in DPD units.

\subsection{Simulation Setup}
\begin{figure}[bht]
\centering
\includegraphics[width=0.25\textwidth]{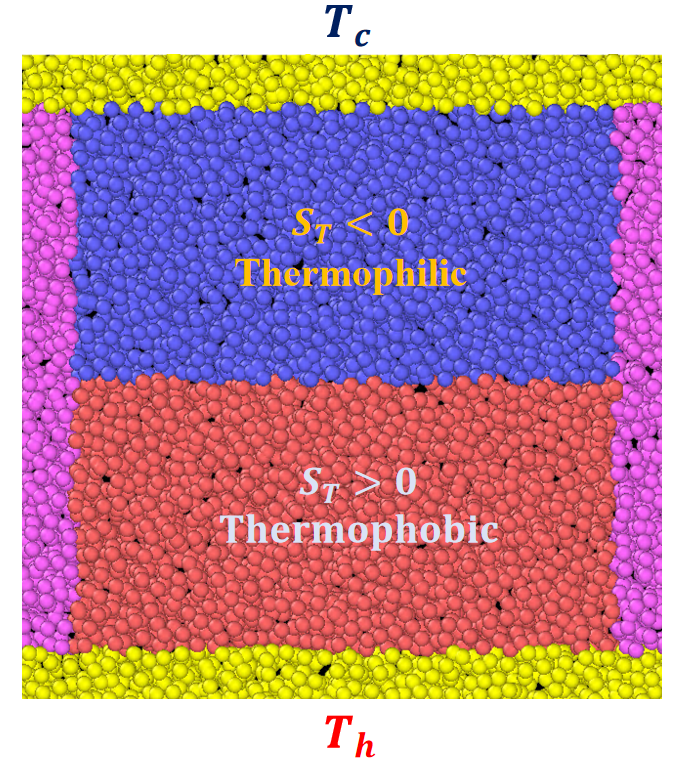}
	\caption{Schematic of a confined square container within thermophilic (blue) and thermophobic (red) particles. }
	\label{img:Sch}
\end{figure}

The simulation setup in this study is illustrated in Fig. \ref{img:Sch}. A confined square container (internal dimensions: $20\times20$ units) with 2-unit-wide sidewalls contains two types of fluid particles exhibiting thermophilic ($S_T<0$) and thermophobic ($S_T>0$) behaviors. 
To model this quasi-two-dimensional system, the thickness of the simulation domain is specified as $4$ units.

When heated from the bottom, both fluid particles experience buoyancy forces within the range of $g \beta \theta \in [0,0.8]$.
Due to the Soret effect, these fluid particles undergo separation, whose extent is governed by the Soret-Rayleigh  number varying within $S_{R,d}\in [0.1,0.4]$. 

In this study, three types of dissipative particle dynamics (DPD) particles were employed: component $A$, component $B$, and fixed boundary particles $W$. The interaction parameters among these particles are summarized in the table \ref{tb:para}. It is worth noting that we use a larger $\kappa C_v$ for fluid and boundary to ensure the temperature continuity near the boundary.

\begin{table}[h]
\caption{\label{tb:para}
The simulation coefficients
}
\begin{ruledtabular}
\begin{tabular}{cccccc}
\textrm{pair}&
\textrm{$a_{ij}$}&
\textrm{$\gamma$ }&
\textrm{$s$}&
\textrm{$R_{C}$}\footnote{conservative and dissipative force and heat conduct use the same cut-off radius.}&
\textrm{$\kappa C_v$ }
\\
\colrule
AA/BB/WW & 25 & 4.5 & 2.0 &1.0&$1$\\
AB       & 25 & 4.5 & 2.0 &1.0&$1$\\
AW/BW    & 25 & 4.5& 2.0 & 1.0&$10$\\
\end{tabular}
\end{ruledtabular}
\end{table}

\section{Results}
\subsection{Velocity and temperature profile}
The velocity and temperature profile are investigated for different $g \beta \theta \in [0,0.8]$ and $S_{R,d}\in [0.1,0.4]$. 
The Prandtl number $Pr$ is $4.5$ for the eDPD parameter set, so that the thermal boundary layer is smaller than the kinetic boundary layer.

\begin{figure}[bht]
\centering
\includegraphics[width=0.55\textwidth]{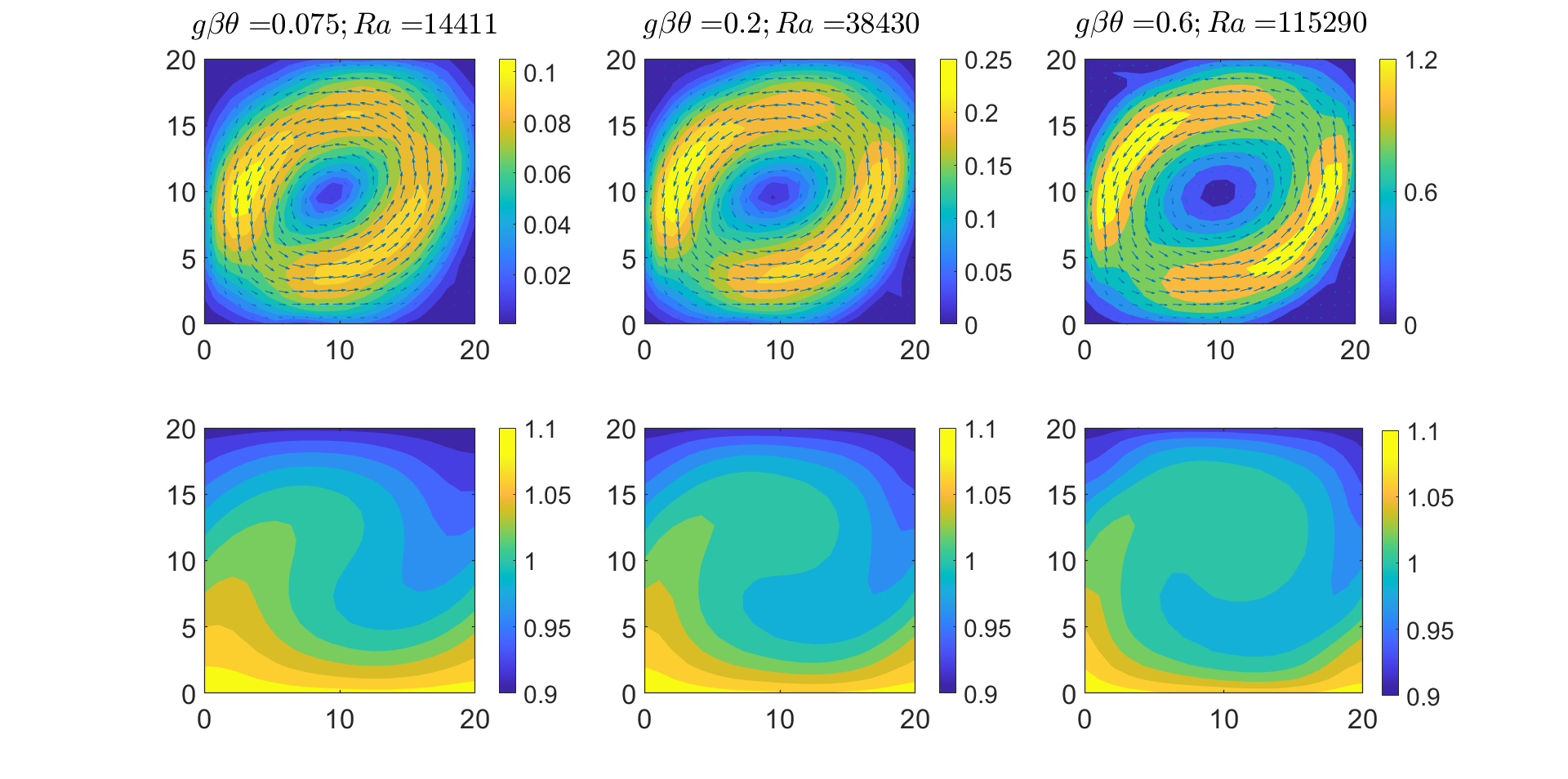}
	\caption{Contours of velocity magnitude, temperature and volume fraction for different buoyancy $g\beta\theta$.}
	\label{img:Contour_for_all_vary_Ra}
\end{figure}

As illustrated in the Fig. \ref{img:Contour_for_all_vary_Ra}, scalar profiles of velocity magnitude, temperature are presented with different temperature difference (Rayleigh number $Ra$) for a certain Soret-Rayleigh number $S_{R,d}=0.3$. 

The increasing temperature difference imparts greater kinetic energy to the flow field, resulting in enhanced maximum flow velocity and elevated Reynolds number. The intensified near-wall flow velocity strengthens convective heat transfer, which manifests as steeper temperature gradients in the near-wall region.

The Reynolds number and Nusselt number is given by
\begin{equation}
Re=\frac{v_0L}{\mu} ; 
Nu=\frac{d\langle T \rangle_B} {dy}\frac{L}{\Delta T}  
\end{equation}
in which $v_0$ is the velocity of a fixed position ($x=5,y=10$) in flow field, ${d\langle T \rangle_B}/{dy}$ is the temprature gardient near the boundary.

\begin{figure}[hbt]
	\centering
\includegraphics[width=0.30\textwidth]{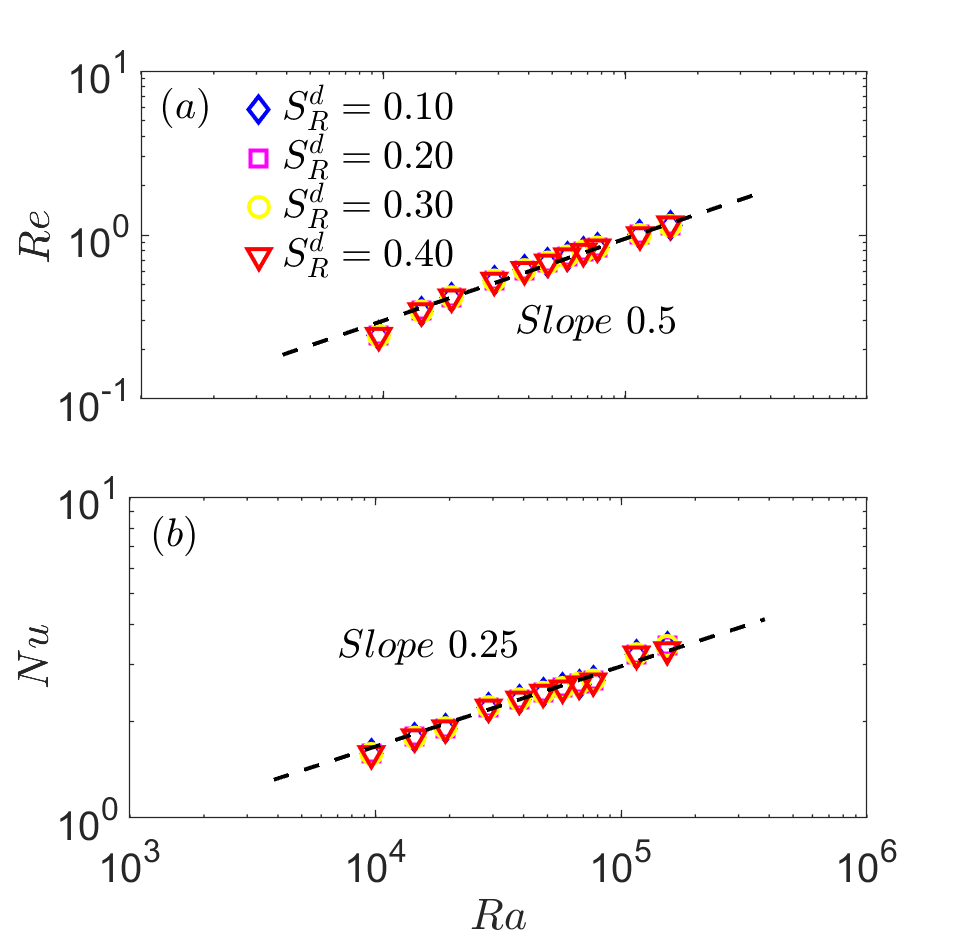}
	\caption{The scaling of $Re$ and $Nu$ with $Ra$.
}
	\label{img:ScalingLaw_ReVsRa_varysT}
\end{figure}

As shown in Fig. \ref{img:ScalingLaw_ReVsRa_varysT}, the  $Re$ and $Nu$ increases with $Ra$ increases and show no distinction for varying $S_R$. This results indicate that the concentration of components is the passive scalar for velocity and temperature.
By fit the results by $Re=Ra^{\zeta}$ and $Nu=Ra^{\gamma}$, we obtain the scaling laws $\zeta = 0.4903$ and $\gamma = 0.2746$. This results can be compared with previous study with large enough Prandtl  number\cite{grossmann2000scaling,shishkina2016momentum}, that is $\zeta=1/2$ and $\gamma=1/4$.

\subsection{Concentration profile and the fraction deviation for long-term mixing}
\begin{figure}[htb]
\centering
\includegraphics[width=0.55\textwidth]{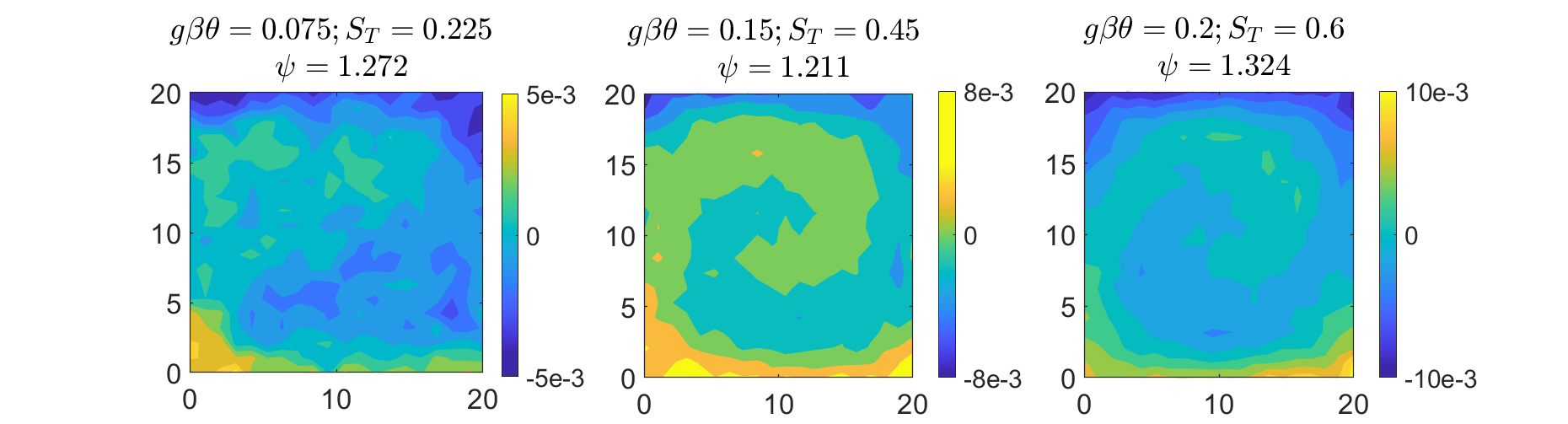}
	\caption{The profile for fraction $\phi-\bar{\phi}$ for different buoyancy $g\beta\theta$ and a fixed $S_R^d=0.3$
}
\label{img:Contour_Line_for_psi}
\end{figure}

After prolonged mixing and diffusion, a slight component separation can still be observed due to the Soret effect and thermal fluctuations, as illustrated in Fig. \ref{img:Contour_Line_for_psi}. To quantitatively characterize this separation, we define the fraction deviation
\begin{equation}
    \psi={\int_0^H\int_0^H |\phi(x,y)-\bar{\phi}|  dx dy};
\end{equation}
The lower fraction deviation represent higher mixing degree.

The calculation of the fraction deviation is inherently dependent on the statistical methodology applied to the concentration field. By divide the flow field into grids of size $\delta x$ and $\delta y$, the average particle number within each grid is statistically quantified over a time interval $T_{aver}$, which given by
\begin{equation}\label{equ:psi_cal}
\psi=\sum_{n_x}\sum_{n_x}\frac{1}{T_{aver}}\int_0^{T_{aver}}(\phi_{t}(n_x\delta x,n_y\delta y)-\bar{\phi})dt;
\end{equation}

Notably, the fraction deviation induced by thermal fluctuations will converge to zero as the averaging time $T_{aver}$ increase. 
To rigorously describe the thermal fluctuation-driven fraction deviation, the averaging time is selected as $T_{aver}=C_v{\delta x^2}/{\alpha}$, in which ${\delta x^2}/{\alpha} \sim 1.0$ is the timescale in DPD units, and the $C_v=1\times 10^5$ is the dimensionless heat capacity\cite{Li2014Energy}.
The timescale enlargement by a factor of 
$C_v$ is required to compensate for the overestimated kinetic energy of fluctuation, as demonstrated in our previous study\cite{Lu2024single}.

\begin{figure}[htb]
\centering
\includegraphics[width=0.45\textwidth]{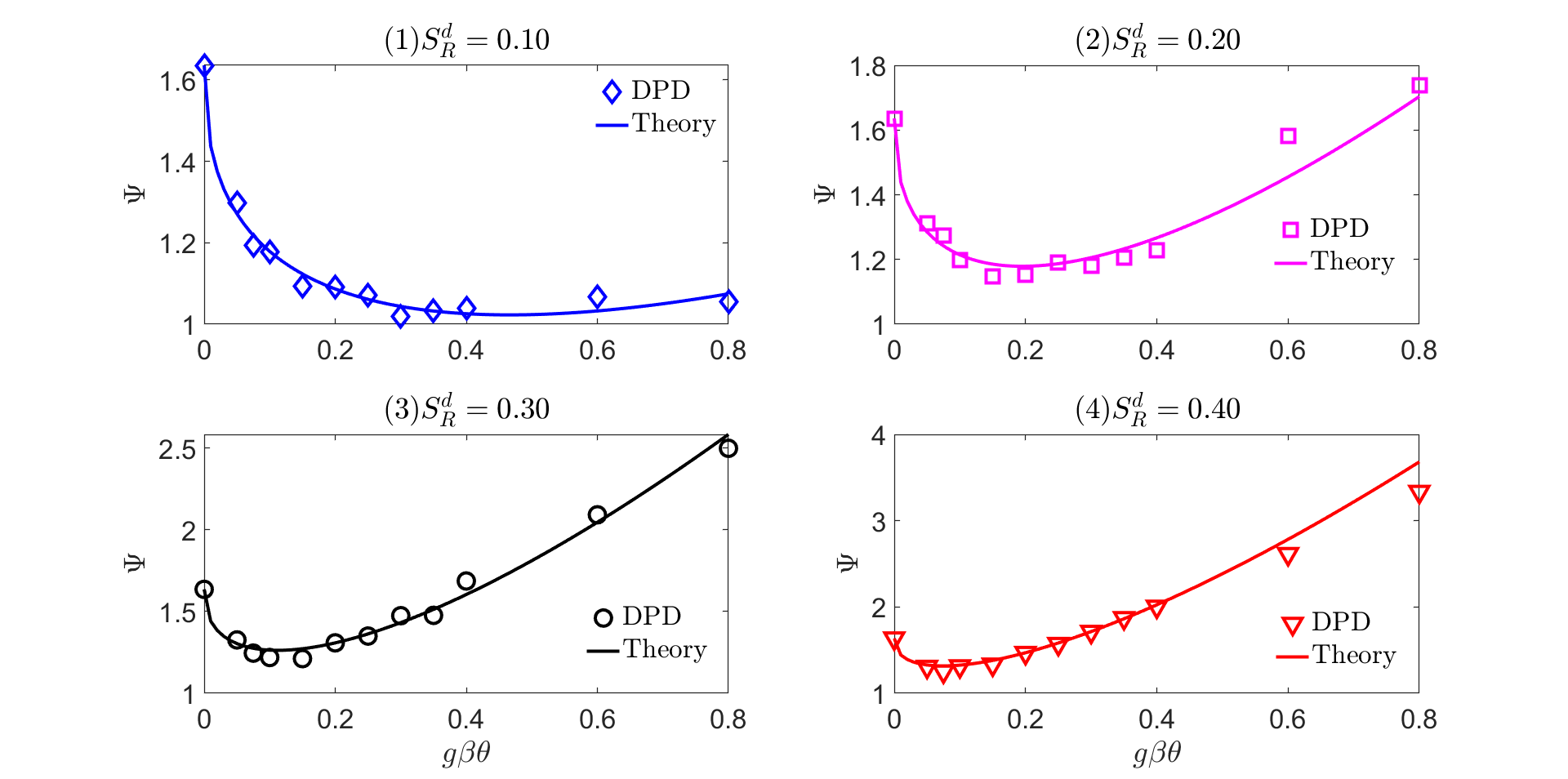}
	\caption{The fraction deviation plot against buoyancy for different Soret-Rayleigh number.
}
\label{img:Psi_vs_gb_vary_S_R}
\end{figure}

Unlike the monotonic enhancement of flow velocity and convective heat transfer with increasing temperature difference, as shown in Fig. \ref{img:Contour_Line_for_psi}, the volume fraction exhibits non-monotonic behavior. 
As shown in Fig. \ref{img:Psi_vs_gb_vary_S_R}, the fraction deviation $\psi$ perform non-monotonic relation with buoyancy $g\beta\theta$.
While the buoyancy-driven convection promotes more uniform components mixing in the central region, the simultaneously increased Soret coefficient $S_T=S_T^{*}\theta^{*}$ induces components separation near the boundary. This counteracting mechanism leads to complex mixing characteristics that demonstrate non-monotonic dependence on temperature difference.

\subsection{Theoretic model}
The fraction deviation $\psi$ have three parts
\begin{equation}\label{equ:theory}
    \psi(Ra,S_R)=\psi_0 -\psi_b +\psi_s;
\end{equation}
in which $\psi_0$ is a constant, which relate to the fluctuation of $\phi$.
$\psi_b$ is the fraction deviation reduction due to convection, which can be obtained by calculate Eq. \ref{equ:psi_cal} when Soret effect is neglected ($S_R^d=0$).
When buoyancy grows, convection become dominate compared with fluctuation, $-\psi_b$ is a monotonic increased function with $Ra$, which can be fitted by $\psi_b \sim Ra^{p_b}$ with fit factor $p_b \approx 0.26$.

$\psi_s$ is the fraction deviation of Soret effect.
Considering the effect of flow filed on mixing, we assume that the fraction deviation of Soret effect has the following format:
\begin{equation}\label{equ:ScaleLaws1}
\psi_s \sim Ra^{p_1}S_R^{p_2}=Ra^{p_1-p_2}S_T^{p_2};
\end{equation}
Using the gradient descent algorithm, factor $p_1=1.34$, $p_2=1.61$. 
It is worth noting that the $p_2-p_1 \approx \gamma \approx 1/4$, if we rearrange Eq. \ref{equ:ScaleLaws1}, we obtain: 
\begin{equation}\label{equ:ScaleLaws1}
\psi_s \sim Ra^{-\gamma}S_T^{1.61};
\end{equation}
where $Ra^{-\gamma}$ can be expressed as $Nu^{-1}$, yielding:
\begin{equation}
\psi_s \sim Nu^{-1}S_T^{1.61}.
\end{equation}
This implies that fraction deviation of Soret effect is proportional to the temperature boundary layers $\lambda_T \sim Nu^{-1}$, that is $\psi_s \sim \lambda_T$.
As shown in Fig. \ref{img:Psi_vs_gb_vary_S_R}, fraction deviation calculated by Eq. \ref{equ:theory} has good agreement with simulation results.

This result can be attributed to the comparison between the temperature field and the concentration field. Both the temperature field and the concentration field exhibit uniformity in the central region, while significant gradients are observed at the boundaries. This implies that noticeable variations in the concentration field occur exclusively within the thermal boundary layer. Furthermore, the relatively large Prandtl number ($P_r>1$) hinders the bulk flow from exerting a substantial influence on the concentration field near the boundaries.
In summary, under conditions of a large Prandtl number, the fractional deviation $\psi_s$ is confined within the thermal boundary layer and exhibits a nonlinear relationship with the Soret coefficient $S_T$.

\section{Conclusion}
In summary, the competition between thermophoretic separation and thermal convective mixing plays a crucial role in multi-component fluid systems driven by buoyancy force. 
We proposed a new dimensionless number, namely Soret-Rayleigh number $S_R=S_T/Ra$, to describe the competition between these two mechanisms, which provides insights into the dynamic balance between Soret effect and thermal convection. 

The fracture deviation $\psi$ was proposed to quantitatively describe the separation index at long time scale.
The findings from energy-conserving dissipative particle dynamics (eDPD) simulations reveal a non-monotonic relationship between the fracture deviation $\psi$ and Rayleigh number $Ra$, highlighting the existence of an optimal $Ra$ that minimized the fracture deviation at long time scales. 
Furthermore, we proposed a theoretical analysis to predict the fracture deviation $\psi$, which successfully describe the non-monotonic behavior.

While our simplified model specifically addresses steady-state conditions at zero separation ratio $\chi=0$, the identified competition mechanism offers qualitative insights applicable to systems with non-zero separation ratios and transient regimes.
These results could provide valuable guidance for optimizing micro-mixer designs and improving their practical applications in various fields.

\begin{acknowledgments}
This research is supported by the Natural Science Foundation
of China (Nos. 12332016, 11832017 and 12172209).

The authors would like to express sincere gratitude to Yi Zhou from Shanghai University for his valuable assistance in the scaling laws analysis of flow and temperature fields. We are also deeply grateful to Weijia Su at Nantong University for his insightful discussions and academic support throughout this research endeavor.
\end{acknowledgments}

\section*{Data Availability}
The data that support the findings of this study are available from the corresponding author upon reasonable request.
\bibliographystyle{unsrt}
\bibliography{aipsamp}

\end{document}